\documentclass[12pt]{article}
\def\beq{\begin{eqnarray}}
\def\eeq{\end{eqnarray}}
\def\beqs{\begin{eqnarray*}}
\def\eeqs{\end{eqnarray*}}
\def\dl{\delta}

\newcommand{{\SD}}{\rm SD}

\newcommand{\lll}{\langle}
\newcommand{\rrr}{\rangle}

\newcommand{\be}{\begin{equation}}
\newcommand{\ee}{\end{equation}}
\newcommand{\T}{\mbox{Tr}\> }

\def\centeron#1#2{{\setbox0=\hbox{#1}\setbox1=\hbox{#2}\ifdim
\wd1>\wd0\kern.5\wd1\kern-.5\wd0\fi
\copy0\kern-.5\wd0\kern-.5\wd1\copy1\ifdim\wd0>\wd1
\kern.5\wd0\kern-.5\wd1\fi}}
\def\ltap{\;\centeron{\raise.35ex\hbox{$<$}}{\lower.65ex\hbox{$\sim$}}\;}
\def\gtap{\;\centeron{\raise.35ex\hbox{$>$}}{\lower.65ex\hbox{$\sim$}}\;}

\begin{document}
\begin{titlepage}

\vskip 1.2cm

\begin{center}

{\LARGE\bf Chiral symmetry breaking in confining theories and
asymptotic limits of operator product expansion}\vskip 1.4cm

{\large  V.I.Shevchenko}
\\
\vskip 0.3cm
{\it State Research Center \\
Institute of Theoretical and Experimental
Physics \\B.Cheremushkinskaya 25, 117218 Moscow, Russia } \\
\vskip 0.25cm e-mail: shevchen@itep.ru

\vskip 2cm

\begin{abstract}
 The pattern of spontaneous chiral symmetry breaking (CSB) in
  confining background
 fields is analyzed. It is explicitly
 demonstrated how to get the inverse square root large proper time
 asymptotic of the operator product expansion which is needed for CSB.
\end{abstract}
\end{center}

\vskip 1.0 cm

\end{titlepage}

\setcounter{footnote}{0} \setcounter{page}{2}
\setcounter{section}{0} \setcounter{subsection}{0}
\setcounter{subsubsection}{0}


\section{Introduction}

The phenomenon of spontaneous chiral symmetry breaking (CSB) is one
of the two most important nonperturbative properties of quantum
chromodynamics (QCD). Another property, confinement of color, is
believed to be deeply connected with CSB. It is interesting to see
how confinement and CSB sometimes play competitive roles when one
speaks about low-energy hadron physics. Namely, one can find
statements in the literature that confinement is "not seen" in
low-energy spectroscopy, and it is possible to model the QCD vacuum
by a set of classical field configurations (as one successfully
does, for example, in the instanton model \cite{cdg}, see reviews
\cite{diak,shuryak} and references therein), despite the fact that
such ensembles typically have no confinement in the sense of area
law of the Wilson loop. There is also an alternative line of
arguments, going back to the good old constituent models, which
states the "supremacy" of confinement (understood in terms of the
confining string formation and corresponding linear potential
between quarks), while CSB is to be derived from confinement (see,
e.g. \cite{simonov}) and there is no need in instantons or alike in
this approach. In fact, these two points of view are to some extent
complementary to each other, as the example of pion clearly shows:
pion is simultaneously Nambu-Goldstone boson which should be
massless in the chiral limit and, on the other hand, it is a bound
state or quarks and antiquarks, as any other meson in QCD is.

The two physical pictures outlined above correspond to two
different approaches one usually uses analyzing the phenomenon of
CSB in QCD. The first approach concentrates on studies of the
Dirac operator zero modes in this or that gauge field background.
 The chiral condensate is related in this case to the (quasi)-zero
 modes density by the well known Banks-Casher relation \cite{bc}. The
model dependence enters when one starts to answer an important
question about physical relevance of the chosen background, in
particular, about its confining properties. We refer the
interested reader to the review \cite{diak} and references therein
for the details of this approach.

Alternatively, one can choose this or that interaction kernel (for
example, manifestly providing confinement) and construct gap
equation for the chiral condensate with this kernel (see, e.g.
recent paper \cite{lesha}). The same procedure is commonly used in
finite density QCD \cite{rajagopal}. This exhibits in a very clear
(but model-dependent) way the general relation between confinement
and CSB \cite{cw}, taking confinement as {\it the cause} of CSB.

In Shifman-Vainshtein-Zakharov sum rules approach \cite{svz} the
chiral condensate $\lll {\bar q} q \rrr$ enters as an input
parameter not directly related to, for example, gluon condensate
$\lll \alpha_s F^2 \rrr$. In other words, there is no simple
relation like $\lll {\bar q} q \rrr = const \cdot \lll \alpha_s F^2
\rrr^{3/4}$ one might naively think of.\footnote{It is worth
reminding that nonperturbative gluon condensate is not a local order
parameter, while chiral condensate is.} In fact, studying
correlators of hadronic currents one can get more sophisticated
relations (see, e.g. \cite{becchi}) between chiral and gluon
condensates (and other nonperturbative quantities like $f_\pi$). The
problem however is that SVZ approach misses the relation between
chiral symmetry breaking and confinement. For example, nonzero value
of $\lll \alpha_s F^2 \rrr$ does not at all indicate that the vacuum
confines.\footnote{As one can see on the lattice at large
temperatures \cite{digi}.} As is well known, one possible way to
take the effects of confinement into account properly is to consider
the dynamics of nonlocal objects like the Wilson loops.

In the present paper we address the problem of CSB in the
first-quantized language, i.e. in terms of quark trajectories and
not fields. In this sense, we are closer to the latter approach
discussed above and not to the former one. The criterium for CSB in
this framework is given by well known Banks-Casher asymptotic law
(eq. (\ref{root}) of this paper). We are going to address the
following question: how
 this asymptotic law follows from the properties of the Wilson
 loop expansion over local condensates and their derivatives.
Speaking differently, we are looking for the simplest subseries of
the Wilson loop operator product expansion, whose summation provides
CSB at the level of one-loop effective action. It is shown that CSB
has to do with the large proper time asymptotic limit of the
specific confining nonlocal  gauge-invariant correlator of gluon
fields.

\section{One-loop effective action}

 We start from the expression for the Euclidean
QCD partition function: $$ Z = \int {\cal D} A_\mu^a{\cal D}q {\cal
D} q^{\dagger} \> \exp\left( -\frac{1}{4g^2}\int d^4x F_{\mu\nu}^a
F_{\mu\nu}^a + \int d^4x q^\dagger \left(i\gamma_\mu D_\mu -
im\right) q \right) =$$ \be = \left\lll \det \left(i\gamma_\mu D_\mu
- im \right) \right\rrr  \ee and we confine our attention to the
case when the mass matrix is proportional to the unit matrix in
flavor space with the eigenvalue $m$. As usual, the average of any
gauge-invariant operator ${\cal R}(A)$ over gauge fields is
understood with the standard Yang-Mills action \be \lll {\cal R}(A)
\rrr = \int {\cal D} A_\mu^a \> {\cal R}(A)\> \exp\left(
-\frac{1}{4g^2}\int d^4x F_{\mu\nu}^a F_{\mu\nu}^a\right)  \ee where
normalization factor, proper gauge-fixing and ghost terms are
included in the integration measure. The chiral condensate is given
by the standard form \be \lll \bar{q} q\rrr^{(M)} =  i \lll
q^{\dagger} q \rrr^{(E)} = - \frac{1}{V} \> \frac{\partial \log
Z}{\partial m} = - \frac{\partial {\Gamma}_{eff}}{\partial
m}\label{cond} \ee and the superscripts $M, E$ stays for Minkowski
and Euclidean values. It is assumed that the right hand side of
(\ref{cond}) does not vanish in the limit $m\to 0$. This nonzero
condensate corresponds to spontaneous CSB.

We are going to exploit the so-called Feynman-Schwinger
representation technique, whose essence goes back to the seminal
papers \cite{f,schw}. There are basically two lines of use of this
approach in the modern research. The first one exploits advantages
the path integration provides for gauge-invariant formulation of
relativistic bound state problems (see review \cite{st} and
references therein). The second line \cite{bk,s} concentrates on the
loop calculations in perturbative field theory (for review see
\cite{schubert}), which is sometimes simpler in world-line approach
than in conventional Feynman diagrammatic framework. We collect some
relevant formulas in Appendix A for convenience, while the
interested reader is referred to the original papers, textbooks
\cite{fh,polyakov,kleinert} and cited reviews for all technical
details.

Confining ourselves by quenched approximation (exact in large
$N_c$ limit), we have  $\lll \det K \rrr = \exp \lll \log \det K
\rrr +{\cal O}(N_c^{-2})$ and hence the standard expression for
Euclidean one-loop effective action:\be \Gamma_{eff} = \frac{1}{V}
\log Z = -2 \int_0^\infty \frac{dT}{T} \exp(-m^2 T) \cdot
\left\lll Z[A,T]\right\rrr \label{gamma} \ee where
 \be Z[A,T] = \int {\cal D} x_\mu \int {\cal D} \psi_\mu
\> \exp\left(-S_0\right)\> {\mbox{tr}}\>
{\mbox{P}}\exp\left(i\int\limits_0^T d\tau\left(A_\mu{\dot x}_\mu -
F_{\mu\nu}\psi_\mu\psi_\nu \right)\right) \label{z} \ee with the
free world-line action given by $ S_0 = \int\limits_0^T d\tau
\left(\frac14 {\dot{x}}_\mu^2 + \frac{1}{2} \psi \dot{\psi}\right)
$. We have included the factor $1/V$ in the integration measure
${\cal D} x_\mu $. The factor $2=4\times (1/2)$ in (\ref{gamma})
came from the trace over anticommuting coordinates (i.e. integration
over the fermionic fields with the free action is normalized to
unity in our conventions). All dynamical information is contained in
the double average (over gauge fields and over quark trajectories)
of the spinor Wilson loop \be w(T)\equiv \left\lll Z[A,T]\right\rrr
\ee There are several limiting cases where one can successfully
study the behavior of $w(T)$ or related functions. Of prime
importance is the nonrelativistic limit of large mass $m$, where the
target space for the contours $x_\mu$ becomes effectively
three-dimensional and introducing the einbein fields for dynamical
masses one can systematically explore constituent picture of hadrons
(see, e.g. \cite{st,ddss} and references therein). However for the
trace (and hence for the closed contours) we are discussing at the
moment nonrelativistically suppressed backward-in-time trajectories
are as important as forward-in-time ones and to address this problem
in einbein fields formalism one probably has to use some alternative
methods (see, e.g. \cite{ns}).

The second important case corresponds to the small $T$ asymptotic.
In a way, this is the standard operator product expansion
\cite{wilson}. In context of the theory of gravity it corresponds to
well known Schwinger - DeWitt expansion \cite{dw}. One is to expand
the Wilson loop in powers of fields in this limit. For constant
background fields this is the way one obtains the effective
Lagrangians of Heisenberg-Euler type. The typical term of this
expansion looks like \be \lll D^{k_1} F^{m_1} D^{k_2} F^{m_2} ...
D^{k_p} F^{m_p} \rrr \cdot T^l \ee The leading term represents the
so called heavy quark condensate \cite{svz} \be \lll \bar{q} q \rrr
= -\frac{1}{12 m} \> \left\lll \frac{\alpha_s}{\pi} F_{\mu\nu}^a
F_{\mu\nu}^a \right\rrr \label{hc} \ee (see also \cite{ant}, where
the heavy quark condensate is discussed in the path integral
formalism, including nonzero temperature case). It is clear that
this expansion is unapplicable in the limit of vanishing mass $m$.

The phenomenon of CSB is related to the large proper time
asymptotic of $w(T)$. As it was noticed by Banks and Casher in
their seminal paper \cite{bc}, for spontaneous CSB one should have
\be w(T) \sim \frac{c}{\sqrt{T}} \;\; \mbox{at} \; T\to \infty
\label{root}\ee Indeed, it is easy to see from (\ref{cond}) and
 (\ref{gamma}) that
 \be
\lll \bar{q} q\rrr = - \frac{\partial \Gamma_{eff}}{\partial m}
\sim -4m \int_0^\infty dT \exp(-m^2 T) \frac{c}{\sqrt{T}} = -
4\sqrt{\pi} \cdot c
 \ee
Taking into account that in free case (i.e. without gauge fields)
$w(T)$ is given by \be w_0(T) = (4\pi T)^{-\frac{d}{2}} \label{e11}
\ee one can say that the quark is dynamically forced to move
effectively in $1$ dimension instead of $3+1$ and this is the cause
for CSB in this framework.

There are a few well known cases where the dimensional reduction of
this kind indeed takes place. However, it is worth noticing that the
condition (\ref{root}) is rather restrictive to be incorporated in a
simple way into the standard background field formalism. Indeed, one
can show on general grounds (see \cite{bv1,bv2,bv3} and \cite{bn}
for review) that the typical large $T$ asymptotic of the heat kernel
trace $ \T {\cal K}(T) = \int dx {\cal K}(T,x,x) $, entering the
effective action as \be \Gamma_{1-loop} = \frac12
\int\limits_0^\infty \frac{dT}{T} \T {\cal K}(T) \ee
 in generic {\it fixed} background
field is given by \be \T {\cal K}(T) = \frac{1}{(4\pi
T)^{{d}/{2}}}\> \left(T W_0 + W_1 + \frac{1}{T} W_2 + ... \right)
\label{larget} \ee where nonlocal factors $W_n$ can be expressed as
integrals of the corresponding zero modes. Notice that the expansion
goes in integer powers of $1/T$. We will address this contradiction
between (\ref{larget}) and (\ref{root}) in the Section 4.

Simple illustrative example is Heisenberg - Euler effective
Lagrangian \cite{he} for constant magnetic field \be L =
-\frac{1}{8\pi^2}\int\limits_0^\infty \frac{dT}{T} \exp(-m^2
T)\cdot\left( \frac{eH}{T}\> \mbox{cth}\> eHT  - \frac{1}{T^2} -
\frac{1}{3} (eH)^2 \right) \label{he} \ee The lowest energy level of
the massless fermion from Dirac sea in constant magnetic field is
zero, hence the absence of exponential damping in (\ref{he}),
$\mbox{cth}\> eHT \to 1$ with $T\to\infty$. The factor $1/T  = T
\cdot T^{-4/2}$ in front of the $\mbox{cth}\> eHT$ corresponds to
the fact that in four-dimensional space-time there are two
directions the fermion can move along as a free particle, while the
dynamics in two other directions is confined by the
field.\footnote{For the field ${\bf H} = {\bf e}_z H$, the particle
moves freely in $z$-directions (along the field) and in
$t$-direction (no force acts on the particle in rest).} This is
nothing but the first term in the rhs of (\ref{larget}). On the
other hand, we see something new here with respect to
(\ref{larget}). The leading term at large $T$ is the last term in
the rhs of (\ref{he}), which is ${\cal O}(T^0)$ and not ${\cal
O}(T^{-1})$. This term represents one-loop short distance charge
renormalization and, at the same time, the leading strong-field
(i.e. small mass) logarithmic asymptotic of (\ref{he}): \be L =
\frac{e^2H^2}{24\pi^2}\log \frac{eH}{m^2} \label{df} \ee This
phenomenon of strong field - short distance duality \cite{ritus}
(small $T$ - large $T$ duality in our context) is quite general
(see, e.g. recent discussion in \cite{dunne}) and provides
interesting possibilities for OPE subseries summation (see
discussion below). Thus we see that quantum dynamics (the necessity
to express the answer in terms of renormalized quantities in this
case) can make the result (\ref{larget}) inapplicable.

\section{Effective action at Gaussian level}

The spinor Wilson loop factor $Z[A,T]$ given by (\ref{z}) can be
expanded in powers of fields and derivatives. It is convenient to
use Fock-Schwinger gauge condition, which is a particular case of
the so called generalized contour (or coordinate) gauge \cite{cg}.
The latter is defined in terms of the oriented non-selfintersecting
contour $z_\mu(s)$ as \be A_\mu(z(s)) \frac{\partial
z_\mu(s)}{\partial s} =0 \ee The simplest contour gauge one usually
uses is the Fock-Schwinger gauge condition with $z_\mu(s) =
x^{(0)}_\mu + s(x-x^{(0)})_\mu$. In this gauge $\partial^2 z_\mu
/\partial s^2 = 0$ and thanks to that  one can express the
vector-potential in the following form (compare with (\ref{field})
from Appendix B) \be A^a_\mu(x) = \int\limits_0^1 sds y_\rho
\exp(sy_\sigma D_\sigma)^{ab} F^{b}_{\rho\mu}(x^{(0)}) \label{fs4}
\ee where $y = x - x^{(0)}$ and Latin indices $a,b = 1, ..., N^2 -1$
stay for adjoint color.

It is worth stressing that for contour gauges gauge invariance
corresponds to the contour independence, not just $x^{(0)}$
independence, as is sometimes posed. This situation is analogous to
the covariant $R_\xi$-gauges, where $\xi$-independence is necessary
but not sufficient condition of gauge invariance; in other words one
can easily construct gauge-noninvariant (and hence physically
unobservable) but $\xi$-independent quantity. The contour
independence is restored only in the full Wilson loop, but not at
any given order of the expansion over fields and/or derivatives
(neither at any given order of the covariant perturbation theory
\cite{bv1,bv2}). Practically it means that general expansion of some
nonlocal object like the Wilson loop over local condensates has no
universal coefficients, independent on the choice of contours used
to fix the gauge (see the Appendix B).

With these reservations in mind, we can proceed and expand the
Wilson loop over gluon fields. In quantum case the dynamics is
determined by the  average over quark trajectories $y_\mu(\tau)$
(and over spinor "coordinates" $\psi_\mu(\tau)$) with Gaussian
weight $\exp(-S_0)$ and over the vacuum gluon fields with the
standard Yang-Mills action. Having performed the latter average, the
first two nontrivial terms in the expansion of $Z[A,T]$ take the
form
$$
\lll Z[A,T] \rrr \approx w_0(T) +  \int {\cal D}y_\mu \int {\cal D}
\psi_\mu \exp(-S_0) \int\limits_0^T d \tau_1 \int\limits_0^{\tau_1}
d\tau_2 \int\limits_0^1 ds_1 \int\limits_0^1 ds_2
$$
$$
(s_1{\dot y}_{\mu_1}(\tau_1) y_{\rho_1}(\tau_1) -
\psi_{\mu_1}(\tau_1) \psi_{\rho_1}(\tau_1) \delta(1-s_1))\cdot(s_2
{\dot y}_{\mu_2}(\tau_2) y_{\rho_2}(\tau_2) - \psi_{\mu_2}(\tau_2)
\psi_{\rho_2}(\tau_2) \delta(1-s_2)) \cdot
$$
\be \cdot \left\lll \T \exp(s_2 y(\tau_2) D) F_{\mu_2
\rho_2}(x^{(0)}) \exp(s_{1} y(\tau_{1}) D) F_{\mu_1 \rho_1}(x^{(0)})
\right\rrr \label{hy} \ee The computational technique for such
integrals is well developed \cite{bk,s,schubert,sch2}. The basic
ingredients are the one-dimensional Green's functions on a circle
$G_B(\tau_1, \tau_2)$ and $G_F(\tau_1, \tau_2)$ defined by \be \lll
y_\mu(\tau_1) y_\nu(\tau_2) \rrr_y = - \delta_{\mu\nu} G_B(\tau_1,
\tau_2) = \delta_{\mu\nu} \left( |\tau_1 - \tau_2| - \frac{(\tau_1 -
\tau_2)^2}{T} \right) \ee and \be \lll \psi_\mu(\tau_1)
\psi_\nu(\tau_2) \rrr_{\psi} = \frac{\delta_{\mu\nu}}{2} G_F(\tau_1,
\tau_2) = \frac{\delta_{\mu\nu}}{2} \mbox{sign}(\tau_1 - \tau_2) \ee
The following identity is of special use \be \left\lll
\exp\left(y_\mu(\tau_1)k^{(1)}_\mu\right)
\exp\left(y_\nu(\tau_2)k^{(2)}_\nu\right) \right\rrr_y = \exp\left(
- G_B(\tau_1, \tau_2) k^{(1)}_\mu k^{(2)}_\mu \right) \ee The result
is given by the expressions \be \Gamma^{(2)}_{eff} =  -2
\int_0^\infty \frac{dT}{T} \exp(-m^2 T)w_0(T) \cdot T^2 K(T)
\label{geff} \ee where $w_0(T)$ is defined by (\ref{e11}), while
$K(T)$ reads \be K(T) = \lll\T F_{\mu\nu} (x^{(0)}) {\cal F}(\xi)
F_{\mu\nu}(x^{(0)}) \rrr \label{uy} \ee with the formfactor \be
{\cal F}(\xi) = \int\limits_0^1 du u(1-u) \exp(u(1-u)\xi) \label{ff}
\ee and $\xi =
T{\overrightarrow{D}}_\sigma{\overrightarrow{D}}_\sigma$. The arrows
indicate that the derivatives act on the right. The contraction of
indices in (\ref{uy}) is worth noticing (see \cite{mul} in this
respect). The most nontrivial thing is the exact expression for the
formfactor (\ref{ff}), which was obtained for the first time in
\cite{bv1} for classical backgrounds (and effectively reproduced in
\cite{sch2} using Feynman-Schwinger technique). For large $m$ eqs.
(\ref{geff})-(\ref{ff}) lead to the expression (\ref{hc}).

\section{Large-$T$ asymptotic limit and CSB}

We are to study large-$T$ asymptotic of (\ref{geff}). One has ${\cal
F}(\xi) \sim 1/\xi^2$ as $\xi \to -\infty$. It is easy to show that
for the scalar particle one would have ${\cal F}(\xi) \sim 1/\xi$ as
$\xi \to -\infty$. This important difference corresponds to the fact
that no exponential damping at large $T$ other than $\exp(-m^2 T)$
is possible for fermions in the chiral limit. Naively one has \be
\lim\limits_{T\to\infty} T^2 K(T) = 2 \lll \T F_{\mu\nu}(x^{(0)})
D^{-4} F_{\mu\nu}(x^{(0)}) \rrr \ee This expression is formal,
however, due to infrared divergencies.  We will show that general
asymptotic expansion at large $T$ may contain logarithmic terms of
the form \be T^2 K(T) = c_0 \log \left(\frac{T}{\lambda^2} \right) +
{\cal O} \left(T^{-1}\>{\log T}\right) \;\;\; {\mbox{for }} T \to
\infty \label{form} \ee where $\lambda$ is typical correlation
length and $c_0$ - some dimensionless coefficient. Possible
subleading $T$-independent contribution in the left hand side of
(\ref{form}) is included into the definition of $\lambda$.

Let us show how the behavior (\ref{form}) of $K(T)$ follows. The
asymptotic pattern is controlled by the function \be f(s) = \lll\T
F_{\mu\nu} (x^{(0)}) \exp\left( s
{\overrightarrow{D}}_\sigma{\overrightarrow{D}}_\sigma \right)
F_{\mu\nu}(x^{(0)}) \rrr \label{uy1} \ee which one needs to know
both at large and at small $s$. The small proper time asymptotic of
$f(s)$ is given by the standard OPE: \be f(s) = \lll\T F_{\mu\nu}^2
\rrr + s \lll\T F_{\mu\nu} D^2 F_{\mu\nu} \rrr + {\cal O}(s^2)
\label{fso}\ee As usual in SVZ sum rules, it is assumed by
definition that all perturbative contributions are subtracted from
each term of (\ref{fso}), thus defining the genuine nonperturbative
function (\ref{uy1}). In the framework of Wilson OPE each term in
(\ref{fso}) depends on the dynamical scale $\mu$ (separating
contributions of perturbative coefficient functions and
nonperturbative matrix elements). The subtle question about $\mu$ -
dependence of the function $f(s)$ and the effective action is
somewhat beyond our main line and will not be discussed. Taking more
phenomenological attitude, the reader may think of $f(s)$ as being
computed on nonperturbative field configurations of one's favorite
QCD vacuum ensemble (instantons, dyons, P-vortices etc); this would
correspond to some effective $\mu \approx 1\> GeV$ of the order of
the onset of nonperturbative dynamics.

The Gaussian approximation to the function $f(s)$ defined by
(\ref{uy1}) can be written as (see details in Appendix B) \be f(s)
\to f_2(s) = \frac{1}{(4\pi s)^2}\int d^4 l \exp\left(- l^2
/4s\right) \lll\T F_{\mu\nu} (x^{(0)}) \exp\left( l_\sigma
{\overrightarrow{D}}_\sigma \right) F_{\mu\nu}(x^{(0)}) \rrr
\label{uy2} \ee The correlator in the right hand side (compare with
(\ref{fs4})) is frequently used in the Gaussian stochastic scenario
of confinement (\cite{ds}, see review \cite{ddss} and references
therein). At distances $l$ larger than some typical correlation
length $\tilde\lambda$ this correlator decays
exponentially\footnote{Notice that this length ${\tilde\lambda}$
characterizes the function $D(l^2)$ and, generally speaking,
${\tilde\lambda} \neq \lambda$. On the other hand, it is physically
natural to assume that  ${\tilde\lambda}$ and $\lambda$ are of the
same order of magnitude.} with $l$. Therefore for
$s{\tilde\lambda}^{-2} \gg 1$ the leading asymptotic is given by $
f(s) \sim s^{-2} $.  To be more precise, it is convenient to
parameterize two-point correlator in the standard way \cite{ds}
$$ \lll\T F_{\mu\nu} (x^{(0)}) \exp\left( l_\alpha
{\overrightarrow{D}}_\alpha \right) F_{\rho\sigma}(x^{(0)}) \rrr =
(\dl_{\mu\rho} \dl_{\nu\sigma} - \dl_{\mu\sigma} \dl_{\nu\rho}
)D(l^2) +
$$
\be + \partial_\mu \left[ (l_\rho \dl_{\nu\sigma} - l_\sigma
\dl_{\nu\rho}) D_1(l^2) \right] - \partial_\nu \left[ (l_\rho
\dl_{\mu\sigma} - l_\sigma \dl_{\mu\rho}) D_1(l^2) \right]
\label{gau} \ee where both functions $D(l^2)$ and $D_1(l^2)$
exponentially decay at large distances. Correspondingly, one has \be
f_2(s) = f_D(s) + f_{D_1}(s) \ee with the following asymptotic
limits at large $s$: \be f_D(s) = \frac{\eta}{s^2} + {\cal
O}(s^{-3}) \;\;\; ;\;\;\; f_{D_1}(s) = \frac{\zeta}{s^3} + {\cal
O}(s^{-4}) \label{fdd1}
 \ee The nonperturbative constants $\eta$, $\zeta$ are given
in this approximation by \be \eta = \frac{3}{4} \int\limits_0^\infty
d l^2 \> l^2 D(l^2) \;\;\; ; \;\;\; \zeta = \frac{3}{16}
\int\limits_0^\infty dl^2 \>l^4\> D_1(l^2) \label{25} \ee

The difference between asymptotic limits of $f_D(s)$ and
$f_{D_1}(s)$ is of crucial importance. Indeed, combining
 (\ref{uy}), (\ref{ff}) and (\ref{fdd1}) one gets \be  K(T) \equiv
K_D(T) + K_{D_1}(T) =  \int\limits_0^1 du \> u(1-u) [f_D(u(1-u)T) +
f_{D_1}(u(1-u)T)] \label{tk} \ee and for $T\to \infty$ we have \be
T^2 K_D(T) \sim \log T \;\; , {\mbox{while}} \;\; T^2 K_{D_1}(T)
\sim \frac{\log T}{T}  \ee The latter term being exponentiated
cannot produce $T^{-1/2}$ term and hence the function $D_1(l^2)$
alone gives no spontaneous CSB. In other words, $D(l^2) \equiv 0$
implies unbroken chiral symmetry. On the other hand, it is well
known \cite{ds} that just nonzero $D(l^2)$ is responsible for
confinement, while $D_1(l^2)$ is not.

It is reasonable to expect that the pattern we have discussed at
the level of two-point correlator is general. From higher
correlators one would have $\sim \log T$ terms, which add up to
the coefficient $\eta$, and $\sim \log^n T$ terms coming from the
corresponding reducible part (i.e. the product of lower order
correlators). The exponentiation of the series produces the
desired  power-like behavior of $w(T)$. All terms $\sim T^{-k}
\log T$ are subleading and do not change the leading power.

Of course, we cannot compute from the first principles the
coefficient in front of a general $\log^n T$ term of this series to
make the above arguments quantitative. But confining ourself by
Gaussian approximation we can proceed further. The actual results
for the chiral condensate and other quantities will crucially depend
on the profile of the function $f_2(s)$, which, in its turn, is
determined by the functions $D(l^2)$ and $D_1(l^2)$. Namely, from
(\ref{uy1}) and (\ref{gau}) we have $$ T^2 K_D(T) = \frac34
\int\limits_0^\infty dl^2 l^2 D(l^2) \int\limits_0^1
\frac{du}{u(1-u)} \exp\left(-\frac{l^2}{4Tu(1-u)}\right) =
$$
$$ = \frac32 \int\limits_0^\infty dl^2 l^2 D(l^2)
\exp\left(-\frac{l^2}{2T}\right) K_0\left(\frac{l^2}{2T} \right) =
$$
\be = \frac32 \int\limits_0^\infty dl^2 l^2 D(l^2)\left[
\log\left(\frac{4T}{e^\gamma l^2} \right) + \frac{l^2}{2T} \>
\log\left(\frac{4T}{e^\gamma l^2} \right) + {\cal O} \left(T^{-2}
\log T\right) \right]\label{logar} \ee Thus the leading large-$T$
asymptotic has the form \be T^2 K_D(T) \to 2\eta
\log\left(\frac{T}{\lambda^2} \right) \label{pol7} \ee with the
correlation length $\lambda$ defined by \be \log \lambda^2 =
\frac{\int\limits_0^\infty dl^2 l^2 D(l^2) \log\left(e^\gamma l^2
/{4} \right)}{\int\limits_0^\infty dl^2 l^2 D(l^2)} \label{pokli}\ee
It is worth repeating that this leading logarithmic term is absent
in the deconfinement phase where $D(l^2) = 0$.

To summarize, the leading large-$T$ asymptotics (\ref{form}) of
$\Gamma_{eff}$ in confining background from the Gaussian term reads
\be \Gamma^{(0)} + \Gamma^{(2)}_{eff} = -2 \int_0^\infty
\frac{dT}{T} \exp(-m^2 T)w_0(T) \left( 1 +  2\eta \log \left(
\frac{T}{\lambda^2} \right) + ... \right) \label{geff1} \ee where
both parameters $\eta$ and $\lambda$ are of essentially
nonperturbative origin.

It is again instructive to compare (\ref{geff1}) with the constant
field case. For constant field the (unrenormalized) result
(\ref{he}) can be written as \be \Gamma = -2 \int_0^\infty
\frac{dT}{T} \exp(-m^2 T)w_0(T) \exp\left(\sum\limits_{n=1}^\infty
\kappa_n (eHT)^{2n}\right) \label{poi} \ee where \be \kappa_n = 2
\frac{(-1)^{n+1}}{n} \frac{\zeta(2n)}{\pi^{2n}} \left( 2^{2n-1} -1
\right) \ee The "correlation length" for the constant field is
infinite and in this sense $T$ is never large, the series in $n$
cannot be truncated and at the Gaussian level $K(T) \to const \neq 0
$ for $T\to\infty$. For finite correlation length typical term in
the corresponding expansion does not increase as a power of $T$ and
$K(T) \sim \log T / T^2 \to 0$ as $T\to\infty$. In terms of
(\ref{poi}) the duality \cite{ritus} mentioned above corresponds to
the fact that the {\it leading} large $T$ asymptotic of renormalized
effective action  (i.e. the expression (\ref{df})) is controlled by
the {\it lowest} term of the {\it unrenormalized} expression
(\ref{poi}) (i.e. the coefficient $\kappa_1$).

The leading Gaussian large $T$ behavior in confining background
given by (\ref{geff1}) is much softer that constant field answer
(\ref{poi}). It is clear that the logarithmic term (\ref{form}) as
it is cannot lead to CSB (since we need power-like rise to get
(\ref{root})). But partial summation of such terms can do the job.
The most crucial point is the structure of the series in the right
hand side of (\ref{geff1}). No universal closed form expressions
analogous to (\ref{uy}), (\ref{ff}) for all terms of higher orders
are known (see, however, \cite{bv3} for explicit form-factors of the
third order). On the other hand, one can argue on physical grounds
that terms \be \frac{1}{n!}\left(2\eta \log
\left(\frac{T}{\lambda^2} \right)\right)^n \ee (together with other
ones) should present. Such terms correspond to the factorized part
contribution of the higher order averages $\lll \T FF ... F \rrr$.
The important role played by these factorized averages is known
under the name of vacuum dominance for a long time and it is
successfully used in sum rules approach.  From general analysis of
\cite{bc} we expect that it is the confinement property that causes
the CSB and it is known for a long time that just described
reduction of the gluon ensemble (known as Gaussian approximation in
the context) provides confinement (see review \cite{ddss} and
references therein). Therefore one can hope that we have summed
"many enough" terms to keep CSB. On the other hand, the connected
parts we have omitted physically correspond to the exchanges by
multi-gluon glueballs and gluelumps, which are heavy objects and
hence their contribution is to be suppressed for the low energy
physics (see \cite{yaya} in this respect). Another argument comes
from the abelian dominance picture (see review \cite{poli} and
references therein). Since in the maximal abelian gauge the higher
irreducible correlators are suppressed (because they correspond to
diagonal -- off-diagonal gluon couplings), one gets the same
factorization pattern. It is important that the Gaussian
factorization is to be assumed for the integration over $x_\mu$ as
well (i.e. we speak about some kind of "rainbow" approximation and
no contraction of $y^{(i)}$, $y^{(j)}$ belonging to different
clusters is done).

Summing of this "Gaussian" subseries of the full Wilson loop average
would result in the effective action
 \be
\Gamma_{eff} =  -2 \int_0^\infty \frac{dT}{T} \exp(-m^2 T)w_0(T)
\cdot \left[\exp(T^2 K(T)) + ... \right] \label{geff2} \ee where
dots stay for the contributions of non-Gaussian terms.
 The spontaneous chiral
symmetry breaking condition has the following form in considered
Gaussian approximation: \be \lim\limits_{T\to\infty} T \frac{d}{dT}
T^2 K(T) = \frac{d-1}{2} \label{csb} \ee or, in terms of (\ref{25})
\be
 \int\limits_0^\infty d l^2 \> l^2
D(l^2) = 1 \label{pol9} \ee  in four dimensional space-time. Then
for the condensate one gets \be \lll \bar{q} q \rrr = -\frac{1}{4}
\left(\frac{1}{\sqrt{\pi} \lambda} \right)^3  \label{condensate} \ee
where $\lambda$ is defined by (\ref{pokli}). The condensate vanishes
in the deconfinement phase transition point.

The above result deserves a few comments. Physically, the parameter
$\eta$ is given by some integral moment of the function $D(l^2)$
and, at the first look can be of arbitrary value, while we need
strictly $\eta = 3/4$ to get CSB. In a sense, this is an artefact of
Gaussian approximation. It is worth repeating that higher terms in
the r.h.s. of (\ref{geff1}) bring both $\sim \log T$ terms, shifting
pure Gaussian value of $\eta$, given by  (\ref{25}) to its pure
"geometrical" value (\ref{csb}) and $\sim \log^n T$ terms, adding up
to the exponent. To some extent it resembles well-known effect of
unphysical surface dependence of the Wilson loop average computed in
Gaussian approximation: to get rid of it one has to sum the full
cluster expansion.\footnote{This analogy should not be understood
literally, of course, there is no any special surface in the
discussed problem, since we sum over all trajectories of the light
quark.} Moreover, suppose that the Gaussian asymptotic law for
(\ref{geff2}) would be different from $\log\> T$ law (\ref{pol7}).
The conclusion in this case would be that Gaussian approximation
does not provide CSB and one needs the full series (or some
subseries other than Gaussian) in (\ref{hy}) to get (\ref{root}).
The actual conclusion is different: Gaussian reduction of the
confining vacuum can lead to CSB (because of (\ref{pol7})) if it is
done in a self-consistent (in the sense of (\ref{csb}),
(\ref{pol9})) way.

On the other hand, all dimensionfull quantities in nonperturbative
theory (like condensates, string tension, correlation lengths etc)
are proportional to the corresponding power of $\Lambda_{QCD}$ with
some dimensionless coefficient, unequivocally fixed by the theory.
Neither the ratio of such coefficients is a freely adjustable
parameter. Since we discuss in this paper not the full theory in the
gluon sector but rather its Gaussian reduction (still keeping
confinement), the relation (\ref{pol9}) can be understood as a kind
of self-consistency condition for Gaussian approximation (if we wish
it to provide CSB). The fact that such self-consistent reduction
should introduce relations between condensates and correlation
lengths is well known \cite{we96}. The ultimate reason for that are
the Bianchi identities: correlator of derivatives
$\epsilon_{\rho\mu\nu\sigma}
\partial_\rho F_{\mu\nu}(x,x^{(0)})$ with any operator, for example,
$F_{\alpha\beta}(y,x^{(0)})$  (inversely proportional to some
typical correlation length) is given by the higher correlators of
$F$'s (another term, containing Bianchi form, vanishes). But the
higher correlators factorize to the product of Gaussian ones in the
chosen approximation.

It is clear from the discussion that we were mostly interested in
qualitative pattern of the effect. Nevertheless it may be
interesting to compare the results with the existing phenomenology
of Gaussian stochastic picture of QCD vacuum. Standard way of
representing the lattice data on the nonperturbative function
$D(l^2)$ is \be D(l^2) =
D(0)P\left(l/\tilde\lambda\right)\exp\left(-l / \tilde\lambda
\right) \ee where $P(x)$ is some rational polynomial, normalized by
condition $P(0)=1$. Lattice \cite{digi} data correspond to $
\tilde\lambda = 0.22$ Fm for quenched $SU(3)$ case, while the value
of dimensionless product $D(0){\tilde\lambda}^4$ is numerically
about $(0.15 \pm 0.05)$. Unfortunately, the poor knowledge of the
pre-exponential factor $P(x)$ precludes one to make quantitative
predictions from the expressions (\ref{pol9}), (\ref{condensate}).
However, it is interesting to notice that the simplest ansatz
$P(x)\equiv 1$ does not describe data satisfactory;\footnote{Notice
that this ansatz approaches the point $l=0$ linearly: $D(l^2) - D(0)
\sim \sqrt{l^2}$ instead of the behavior $D(l^2) - D(0) \sim l^2$
dictated by (\ref{uy2})} it predicts $D(0){\tilde\lambda}^4 = 1/12$
and too small value of the chiral condensate $\lll \bar{q} q \rrr =
-(140 \>\mbox{MeV})^3$ instead of the correct value $\lll \bar{q} q
\rrr = -(250 \>\mbox{MeV})^3$. Of course, one can easily fit $P(x)$
to get the desired numbers. But this seems to be misleading since,
as it has been already mentioned, eventually these are the higher
order terms which shift $\eta$ and $\lll \bar{q} q \rrr$ to their
correct values.

\section{Conclusion}

We have studied the chiral symmetry breaking at one-loop level in
the background of confining gluon fields. The latter is taken in the
Gaussian picture, i.e. we have omitted higher than two-point
irreducible condensates. This approximation is supported by the
sum-rule phenomenology as well as by more sophisticated analytical
and lattice analysis. It is worth stressing that the vacuum gluon
fields ensemble reduced in such way still has the confinement
property. It is shown that this vacuum breaks chiral symmetry
spontaneously provided the large proper-time asymptotic of the
operator product expansion has the form (\ref{csb}).

At the same time it is worth stressing that despite "Gaussian
reduction" alone is enough to get CSB, it does not mean that higher
order terms are not important. There are at least two respects where
they are: first, their presence is a matter of principle for
correctness of the general Wilson loop asymptotic (see, e.g.
comments in \cite{ddss}) like surface independence etc; and second,
they can be numerically significant (say, contribute 10-15\%) in the
quantities like $\eta$, $\zeta$.

We have also discussed that perturbative asymptotic of the
corresponding correlators cannot provide CSB; roughly speaking, at
small momenta perturbative Green's functions are "too soft". No any
specific topological properties of the vacuum gluon fields show up
in our analysis, the only relevant thing is the large proper time
asymptotic of the OPE (resulting from confinement). We did not use
in any place explicit expressions for the gluon fields profile, only
the vacuum averages entered our analysis. Nevertheless it would be
interesting to understand the relation of the discussed issues with
the well established phenomenology of CSB in, e.g. instanton
backgrounds (without confinement {\it per se}), having in mind that
it is ultimately the strong confining color forces that determine
the dynamics of CSB in accordance with general analysis of
\cite{cw}.

\bigskip

{\bf Acknowledgments }

\medskip

The author expresses his gratitude to the "Dynasty" foundation and
ICFPM for financial support. The support from the grant
MK-1530.2005.2 and from the Federal program of the Ministry of
industry, science and technology No.40.052.1.1.1112 is also
acknowledged. Stimulating discussions with Yu. Simonov and
V.Zakharov are acknowledged.

\bigskip

 {\it Note added}

The related issues are discussed in the recent preprint
hep-th/0511176 by V.Vyas we got after our paper had been completed.
The main interest of the cited paper concentrates on the
consequences the asymptotic law (\ref{root}) has for the behavior of
two-point correlators of hadronic currents, while we have tried to
answer another and physically different question: how can the
asymptotic (\ref{root}) itself be obtained using the OPE language.

\newpage

{\bf\large Appendix A}

Throughout the paper we use Feynman-Schwinger world-line
integration technique.
The standard building blocks here are the following: \\
1) $\gamma_5$ - invariance of the determinant (we define $D_\mu =
\partial_\mu - i A_\mu$):
$$ \log \det \left(\gamma_\mu
D_\mu + m \right) = \frac12 \log \det \left(\left[\gamma_\mu D_\mu
+ m \right] \left[- \gamma_\mu D_\mu + m \right]\right)=
$$
\be = \frac12 \log \det \left( - D^2  + \frac{ig}{4} F_{\mu\nu}
\left[\gamma_\mu \gamma_\nu \right] + m^2\right) \ee  2)
proper-time representation of the logarithm: \be \log A = -
\int\limits_0^{\infty} \frac{dT}{T} \exp\left(-AT\right)
\label{log}\ee The above expression is of course symbolical and
properly regularized form of (\ref{log}) must be used in actual
computations. The typical examples are given by exact equality \be
\log A + \gamma = - \lim\limits_{\xi \to 0} \int\limits_0^{\infty}
dT T^{\xi-1} \left(1+\xi\log T\right)\exp\left(-AT\right) \ee
which can be obtained differentiating the identity $A^{-\xi}
\Gamma(1+\xi) = \xi \int_0^\infty dT T^{\xi -1} \exp(-AT)$ or by
frequently used integral \be \log A = - \int\limits_0^\infty
\frac{dT}{T} (\exp(-AT) - \exp(-T)) \ee

3) integration over commuting paths replacing the bosonic traces:
$$ \T \exp\left(-T(-D^2)\right) = \int \frac{d^4 p}{(2\pi)^4} \>
\lll p | \exp\left(-(p+gA)^2 T\right) |p\rrr =
$$
\be = \int {\cal D} x \exp\left(-\frac14 \int\limits_0^T
{\dot{x}}_\mu^2 d\tau \right) \> {\mbox{tr}}\>{\mbox{P}}\exp
\left(ig\int\limits_0^T A_\mu(x) {\dot{x}}_\mu d\tau\right)\ee 4)
integration over anti-commuting fields representing the path-ordered
exponent of the gamma matrices: \be \T {\mbox P}\exp \left( -
\frac{ig}{4} \int\limits_0^T d\tau F_{\mu\nu} \left[\gamma_\mu
\gamma_\nu \right] \right) = \int {\cal D} \psi \> {\mbox{tr}}\>
{\mbox{P}}\exp\left( - \int\limits_0^T d\tau \left(\frac{1}{2}
\psi_\mu {\dot{\psi}}_\mu - ig F_{\mu\nu} \psi_\mu \psi_\nu
\right)\right) \ee where the small trace ${\mbox{tr}}$ goes over
color indices only.  The corresponding operators $\hat\psi$
anti-commute and act on the Hilbert space of Dirac spinors
$|\alpha\rrr$ as \be \psi_\mu\psi_\nu + \psi_\nu \psi_\mu =
\delta_{\mu\nu} \;\; ; \;\; \psi^\mu |\alpha\rrr =
\frac{1}{\sqrt{2}}\gamma^\mu_{\alpha\beta} |\beta \rrr \ee The
normalization is provided by \be \int {\cal D} \psi \exp\left( -
\frac12 \int\limits_0^T d\tau \psi_\mu {\dot{\psi}}_\mu \right) = 1
\ee  The integration goes over periodic commuting $x_\mu(\tau)$ and
anti-commuting $\psi_\mu(\tau)$, defined on the circle of the length
$T$.

\bigskip

{\bf \large Appendix B}

Let us consider the gauge invariant phase factor along small closed
contour $C$ for a particular inhomogeneous field $A_\mu$ \be W(C) =
\Phi(x,x) = \T \mbox{P}\exp\left(i\oint\limits_C A_\mu dz_\mu
\right) \ee The naive expansion over fields has the following form:
\be W(C) \approx 1 + i^2 \oint\limits_C dz_\mu \oint\limits_C^z
du_\nu \> \T A_\nu(u) A_\mu(z) + ... \label{wilson} \ee The second
term is gauge non- invariant (in nonabelian case). If one rewrites
it in contour gauge \cite{cg} where \be A_\mu(x) = \int\limits_0^1
ds \frac{\partial z_\alpha(s)}{\partial s}\frac{\partial
z_\beta(s)}{\partial x_\mu}\>F_{\alpha\beta}(z(s)) \label{field} \ee
the answer would be contour dependent. One can rearrange the series
(\ref{wilson}) to make it manifestly gauge-invariant: \be W(C)
\approx 1 + i^2 \int\limits_S d\sigma_{\mu\rho}(z) \int\limits_{S_z}
d\sigma_{\nu\sigma}(u) \> \T F_{\mu\rho}(z,x^{(0)})
F_{\nu\sigma}(u,x^{(0)}) \label{ei} \ee where $x^{(0)} = z(s=0)$ and
the shifted field strength is given by \be F_{\mu\rho}(z,x^{(0)}) =
\Phi(x^{(0)}, z) F_{\mu\rho}(z) \Phi(z, x^{(0)}) \ee This is nothing
but the nonabelian Stokes theorem \cite{cg,nast}. Notice that
despite the Wilson loop itself is gauge-invariant and
$x^{(0)}$-independent (it depends only on the contour $C$ but not on
the integration surface $S$ in (\ref{ei})), the second term in the
right hand side of (\ref{ei}) is gauge-invariant {\it but} contour
dependent. Its SVZ-like expansion over condensate and derivatives
would manifestly depend on the particular choice of contours
$z_\mu(s)$ and will be {\it different} for different contour gauges
(i.e. for different choices of $z_\alpha(s)$ and hence $S$).

However, in Gaussian approximation the account for contour
dependence would not be legitimate \cite{ds}. The reason is that the
variation of the nonlocal Gaussian correlator
$$
\lll \T F_{\mu\rho}(z,x^{(0)})F_{\nu\sigma}(w,x^{(0)}) \rrr
$$
over contours brings additional powers of $F_{\alpha\beta}$, and
hence it is proportional to the correlators of higher orders (see
related discussion in \cite{ddss}). We have used the same phenomenon
replacing (\ref{uy1}) by (\ref{uy2}) in the main text. Indeed, let
us look at the power expansions of those functions. They are given
by \be f(s) = \lll\T F_{\mu\nu} \left( 1+  s D_\alpha^2 +
\frac{s^2}{2} (D_\alpha^2)^2 + ... \right) F_{\mu\nu} \rrr \ee and
\be f_2(s) = \lll\T F_{\mu\nu}  \left( 1+  s D_\alpha^2 +
\frac{s^2}{2} \left[(D_\alpha^2)^2 -\frac23 D_\alpha
iF_{\alpha\beta} D_\beta + \frac16 F_{\alpha\beta}^2 \right] + ...
\right) F_{\mu\nu} \rrr \ee It is seen that the difference between
$f(s)$ and $f_2(s)$ contains correlators of higher orders, resulting
from non-commutativity of $D_\mu$'s. In Gaussian approximation each
power of derivative corresponds to large  factor
${\tilde\lambda}^{-1}$ and in this sense to extract the leading term
one can always neglect the commutators $[D_\alpha D_\beta]$. The
validity of Gaussian approximation, in its turn, is controlled by
the dimensionless parameter
$$
\kappa = \lll \T F_{\mu\nu}^2 \rrr {\tilde\lambda}^4
$$
which is assumed to be small (it was mentioned above that, for
example, $D(0){\tilde\lambda}^4$ is about $1/10$ according to the
lattice data). In physical terms, Gaussian vacuum is
short-correlated, and the nonperturbative gluon condensate is small
in units of the lightest gluelump mass (the latter is of the order
of $1/\tilde\lambda$). Using the language of covariant perturbation
theory one can say that we have summed up only terms of the kind
$\lll \T F (D^2)^n F \rrr$ and not the terms with lesser number of
derivatives.

To summarize, we have
$$
f(s) = f_2(s) + {\cal O}(\kappa^{1+r})
$$
where $r$ is some positive number. Therefore the leading Gaussian
asymptotic of $f(s)$ is given by the asymptotic of $f_2(s)$ found in
eq. (\ref{fdd1}) in the main text.

\newpage


\end{document}